\documentclass[aps,showpacs,prb,twocolumn,superscriptaddress,floatfix]{revtex4-2}

\usepackage{graphicx,bm,amssymb,amsmath,dcolumn,hyperref}
\usepackage{multirow}
\usepackage[shortlabels]{enumitem}
\usepackage{color}
\usepackage{subfigure}  
\usepackage{epstopdf}
\usepackage{bbm}
\usepackage{graphicx}
\usepackage{hyperref}
\usepackage{threeparttable}
\usepackage{amsthm}
\usepackage{mathtools}
\usepackage{color}
\usepackage{tikz}
\usepackage{float}
\usepackage{empheq}
\usepackage{algorithm}
\usepackage{algpseudocode}
\usepackage{array}
\usepackage{tabularx}

\begin{document}

\title{Quantum Path-integral Method for Fictitious Particle Hubbard Model}

\author{Zhijie Fan}
\email{zfanac@ustc.edu.cn}
\affiliation{Hefei National Research Center for Physical Sciences at the Microscale and School of Physical Sciences, University of Science and Technology of China, Hefei 230026, China}
\affiliation{Hefei National Laboratory, University of Science and Technology of China, Hefei 230088, China}
\affiliation{Shanghai Research Center for Quantum Science and CAS Center for Excellence in Quantum Information and Quantum Physics, University of Science and Technology of China, Shanghai 201315, China}

\author{Tianning Xiao}
\affiliation{Hefei National Research Center for Physical Sciences at the Microscale and School of Physical Sciences, University of Science and Technology of China, Hefei 230026, China}

\author{Youjin Deng}
\email{yjdeng@ustc.edu.cn}
\affiliation{Hefei National Research Center for Physical Sciences at the Microscale and School of Physical Sciences, University of Science and Technology of China, Hefei 230026, China}
\affiliation{Hefei National Laboratory, University of Science and Technology of China, Hefei 230088, China}
\affiliation{Shanghai Research Center for Quantum Science and CAS Center for Excellence in Quantum Information and Quantum Physics, University of Science and Technology of China, Shanghai 201315, China}

\date{\today}

\begin{abstract}
We formulate a path-integral Monte Carlo algorithm for simulating lattice systems consisting of fictitious particles governed by a generalized exchange statistics. This method, initially proposed for continuum systems, introduces a continuous parameter $\xi$ in the partition function that interpolates between bosonic ($\xi = 1$) and fermionic ($\xi = -1$) statistics. We generalize this approach to discrete lattice models and apply it to the two-dimensional Hubbard model of fictitious particles, including the Bose- and Fermi-Hubbard models as special cases. By combining reweighting and $\xi$-extrapolation techniques, we access both half-filled and doped regimes. In particular, we demonstrate that the method remains effective even in strongly correlated, doped systems where the fermion sign problem hinders conventional quantum Monte Carlo approaches. Our results validate the applicability of the fictitious particle framework on lattice models and establish it as a promising tool for sign-problem mitigation in strongly interacting fermionic systems.
\end{abstract}

\maketitle

\section{Introduction}

The Quantum Monte Carlo (QMC) method is one of the most powerful and versatile numerical techniques for investigating quantum many-body systems, with broad applications across diverse areas of physics~\cite{foulkes_quantum_2001,sandvik2010,carlson2015}. By reformulating the partition function as a weighted sum over configurations in a chosen computational basis, QMC employs importance sampling to explore the exponentially large configuration space efficiently. However, when applied to fermionic systems, QMC suffers from the notorious ``fermion sign problem"~\cite{loh1990}: configuration weights oscillate in sign due to the antisymmetric nature of fermionic wavefunctions, leading to severe cancellations and overwhelming statistical noise. As a result, the computational cost required to achieve a certain accuracy grows exponentially with both the system size and inverse temperature, which severely limits the applicability of QMC to many important fermionic systems~\cite{troyer2005}. 

A prominent example is the Fermi-Hubbard model~\cite{hubbard1963}, a paradigmatic strongly correlated lattice fermion system that is closely related to the mechanism of high-temperature superconductivity and the origin of quantum magnetization~\cite{keimer2015, arovas2022}. Over the past 60 years, remarkable progress has been made in numerical~\cite{leblanc2015, qin2022}, and experimental~\cite{bohrdt_exploration_2021, shao2024} studies of this model. Yet, obtaining unbiased solutions of the Hubbard model at generic fillings and interaction strengths remains an outstanding challenge. The determinant quantum Monte Carlo (DQMC) method, widely used for exploring the equilibrium properties of interacting fermions~\cite{blankenbecler1981,hirsch1983,white1989,assaad2008}, is sign-problem-free only at half-filling on bipartite lattices due to particle-hole symmetry. Away from this special point, the sign problem reemerges and worsens rapidly, severely limiting the practical applicability of DQMC~\cite{white1989}. Furthermore, in the strongly interacting regime, which hosts rich and intriguing phenomena, such as Nagaoka ferromagnetism~\cite{nagaoka1966, thouless_exchange_1965, tasaki1989}, DQMC suffers not only from sign problems but also from severe numerical instabilities and ergodicity issues --- even at half-filling where the sign problem is absent~\cite{khatami_finite-temperature_2015}.

The fermion sign problem has been proven to be NP-hard (nondeterministic polynomial-time hard)~\cite{troyer2005}, implying that finding a generic solution to it is extremely difficult, if not impossible. Despite this inherent difficulty, significant progress has been made in developing approaches that alleviate or avoid the sign problem in specific models or parameter regimes~\cite{li2019, PAN2024879}, including the diagrammatic QMC~\cite{kozik_diagrammatic_2010}, the Lefschetz thimble method~\cite{ulybyshev2017, ulybyshev_lefschetz_2020}, the Majorana representation QMC~\cite{li_solving_2015}, the constrained-path QMC~\cite{he2019}, and the sign bound theory~\cite{zhang_fermion_2022}. These developments underscore the need for new conceptual frameworks to broaden the applicability of QMC to fermionic systems.

Very recently, a novel approach to circumvent the fermion sign problem based on the path-integral Monte Carlo (PIMC) simulations of fictitious particles was proposed by Xiong and Xiong~\cite{xiong_thermodynamic_2022, xiong2022, xiong2024}. In this framework, the canonical partition function is parametrized by a continuous parameter $\xi \in \mathbb{R}$, which defines a family of fictitious identical particles with generalized exchange statistics interpolating between Bose–Einstein $(\xi=1)$ and Fermi–Dirac $(\xi=-1)$ limits. By carrying out PIMC simulations of fictitious particles in the $\xi>0$ sector, where all configuration weights remain strictly positive, thus avoiding the fermion sign problem, one can then extrapolate the physical observables to the fermionic limit of $\xi = -1$~\cite{xiong_thermodynamic_2022,xiong2022,xiong2024}.

The fictitious particle PIMC approach has achieved notable success in simulating interacting fermion systems in continuous space. Early studies proposed a $\xi$-extrapolation method to the interacting fermion gas where the energy of the system can be accurately obtained using an empirical quadratic fit in $\xi$~\cite{xiong_thermodynamic_2022}. Subsequent work by Dornheim et al. demonstrates that this method yields excellent agreement with benchmark results for various observables in weakly degenerate systems~\cite{dornheim_fermionic_2023}. In particular, for the warm dense uniform electron gas, they report a speedup exceeding $11$ orders of magnitude over direct fermion PIMC simulations. In a recent study, Dornheim et al.~\cite{dornheim2024a} employed the fictitious particle PIMC to perform highly accurate simulations of the warm dense electron gas, achieving system sizes of up to $\sim 1000$ particles. Although the $\xi$-extrapolation becomes ineffective at low temperatures, a refined extrapolation strategy guided by the physical properties of the fictitious particles has been developed to study the thermodynamic properties of fermions at arbitrary temperatures~\cite{xiong2023}. The method has been successfully employed to study the normal liquid state of Helium-3, a system with high quantum degeneracy, providing energy values that match experimental measurements~\cite{morresi_normal_2025}. Despite these successes, the application of the fictitious particle framework has thus far been restricted to continuous-space models. Extending this approach to lattice systems represents a compelling direction for future research, offering the possibility of alleviating the fermion sign problem in a broader class of models and accessing parameter regimes that remain challenging for conventional QMC methods.

In this work, we extend the fictitious particle framework to discrete lattice models and develop a PIMC algorithm for fictitious identical particles on a lattice. Specifically, we apply the algorithm to the two-dimensional Hubbard model of fictitious particles and evaluate its effectiveness for investigating the corresponding fermionic model. We implement a $\xi$-based reweighting scheme and assess its applicability to both half-filled and doped cases, showing that this seemingly simple approach remains effective in the strong-coupling regime. Furthermore, we show that $\xi$-extrapolation further broadens the applicability of reweighting. Together, these techniques enable access to strongly interacting, doped Hubbard models that are otherwise challenging for conventional QMC methods.

The paper is organized as follows: Section~\ref{section:fictitious_particle_partition_function} gives a brief introduction to fictitious particle formalism. In Section~\ref{section:algorithm}, we formulate the path-integral representation of fictitious particles on a lattice and present the corresponding PIMC algorithm. Section~\ref{section:hubbard_model} applies the fictitious particle PIMC method to the two-dimensional Hubbard model to evaluate the performance of the method. Section~\ref{section:extrapolation} demonstrates the application of the $\xi$-extrapolation technique. Finally, in Section~\ref{section:discussion}, we discuss the broader applicability, limitations, and future directions of the method.

\section{Partition Function of Fictitious Identical Particles}
\label{section:fictitious_particle_partition_function}
The quantum statistical partition function for a system characterized by Hamiltonian $\mathcal{H}$ is defined as:
\begin{align}
    \mathcal{Z} &= \mathrm{Tr}\left[e^{-\beta \mathcal{H}} \right]\\
    &= \sum_{\alpha} \left\langle \Psi_\alpha \right|e^{-\beta \mathcal{H}}\left| \Psi_\alpha \right\rangle,
\end{align}
where $\beta\equiv1/T$ is the inverse temperature, and $\{\left| \Psi_\alpha \right\rangle\}$ represents an orthonormal and complete basis in the Hilbert space of $\mathcal{H}$. For a system of $N$ distinguishable particles, each having $M$ possible single-particle states, the Hilbert space dimension is $M^N$. A basis vector can be represented as:
\begin{align}
    \left| \Psi_\alpha \right\rangle = \left| \psi_1,\psi_2,\cdots,\psi_N \right\rangle,
\end{align}
where $\psi_i$ labels the state of the $i$th particle in a chosen single-particle basis, e.g., spatial position, spin. An arbitrary wave function in this space is a linear combination of basis vectors $\left|\Phi\right\rangle = \sum_\alpha c_\alpha\left|\Psi_\alpha\right\rangle$ with normalized coefficients satisfying $\sum_\alpha |c_\alpha|^2 = 1$. For identical particles, the many-body wave function must also be (anti)symmetric under particle exchange: symmetric for bosons and antisymmetric for fermions. Accordingly, the Hilbert space can be constructed using basis states that respect the required exchange symmetry. For bosons, a natural choice is the occupation number basis, while for fermions, the basis states can be constructed using Slater determinants to ensure antisymmetry.

An alternative way to express the canonical partition function of $N$ identical particles is to incorporate the effects of particle exchanges explicitly,
\begin{align}
    \mathcal{Z}_{\pm}(T) = \frac{1}{N!}\sum_{P \in S_N} \sum_{\alpha} (\pm1)^{N_P}\left\langle \Psi_\alpha \right|e^{-\beta \mathcal{H}}\left| P(\Psi_\alpha) \right\rangle,
    \label{eq:partition_function_fermi_boson}
\end{align}
where the factor $-1$ corresponds to fermions and $+1$ to bosons~\cite{xiong2022,xiong_thermodynamic_2022}. The first summation goes over all permutations in the $S_N$ permutation group, which consists of $N!$ elements. Here, $N_P$ is the minimal number of pairwise swaps to recover the original order from permutation $P$. A permutation $P$ acts on a basis $\left|\Psi\right>$ by reassigning particle labels, such that the state of the $i$th particle is mapped to the $P(i)$th particle, resulting in a permuted basis state
\begin{align}
\left|P(\Psi)\right> = \left|\psi_{P(1)}, \psi_{P(2)}, \cdots, \psi_{P(N)} \right>.
\end{align}
For fermions, the summation in the partition function can have negative terms when $N_P$ is odd. This could eventually invalidate the probabilistic interpretation of configuration weights in QMC sampling, leading to the fermion sign problem~\cite{li2019, ying2019}.

To alleviate the sign problem, Ref~\cite{xiong_thermodynamic_2022} proposes a generalized partition function,
\begin{align}
    \mathcal{Z}_{\xi}(T)= \frac{1}{N!}\sum_{P \in S_N} \sum_{\alpha} \xi^{N_P}\left\langle \Psi_\alpha \right|e^{-\beta \mathcal{H}}\left| P(\Psi_\alpha) \right\rangle,
    \label{eq:generalized_partition_function}
\end{align}
where $\xi$ is a real number, allowing one to continuously interpolate between fermionic ($\xi=-1$) and bosonic ($\xi=1$) statistics. 
Equation~\eqref{eq:generalized_partition_function} describes a system of identical fictitious particles obeying a generalized quantum statistics~\cite{isakov_generalization_1993},
\begin{align}
    a^{\dagger}_i a_j -\xi a_j a_i^{\dagger} = \delta_{ij}.
    \label{eq:generalized_exchange_statistics}
\end{align} When $\xi=+1(-1)$, the bosonic (fermionic) statistics can be recovered. In the bosonic sector $\xi >0$, the negative sign due to particle exchanges is absent. When $|\xi|<1$, the contribution from particle permutations is suppressed; at $\xi=0$, all exchange terms vanish, corresponding to an idealized model of distinguishable quantum particles known as Boltzmannons~\cite{dornheim_fermionic_2023}. The fictitious particle formulation thus provides a framework that smoothly connects bosonic, fermionic, and distinguishable particle limits, and forms the basis for developing a path-integral Monte Carlo algorithm that operates in the $\xi > 0$ regime to circumvent the fermion sign problem. 

\section{Path-integral Method for Fictitious Particles on Lattice}
\label{section:algorithm}
In this section, we formulate the path-integral representation for fictitious particles on lattices, building upon the generalized partition function defined in Eq.~\eqref{eq:generalized_partition_function}. We develop a PIMC algorithm that employs a modified worm update scheme by explicitly incorporating the particle exchange number $N_P$, thereby enabling direct simulation of fictitious particles in the $\xi>0$ sector.
Furthermore, we implement a reweighting scheme in $\xi$ to estimate physical observables in the $\xi < 0$ regime based on simulations performed at $\xi > 0$. Together, the modified worm algorithm and the $\xi$-reweighting scheme provide a framework for studying lattice models with generalized quantum statistics.

\subsection{Path-integral formalism for Fictitious Particles}
The central idea of the PIMC method~\cite{ceperley1989, prokofev_worm_1998, boninsegni2006a, assaad2008} is to map a $d$-dimensional quantum system onto an equivalent $(d+1)$-dimensional classical system by expanding the quantum partition function in a suitable basis. In this mapping, the additional dimension corresponds to the imaginary time ($\tau$) axis with a periodic boundary condition. The quantum statistical properties are then extracted by sampling configurations in the classical system. Each configuration consists of a sequence of states along the imaginary-time axis, forming continuous trajectories in the $(d+1)$-dimensional space, namely the worldlines. For a system with particle number conservation, the trajectories of particles are closed loops. The partition function of the original quantum model is then represented as a weighted sum over all possible worldline configurations in $(d+1)$-dimensional space-time.

\begin{figure}
    \centering
    \includegraphics[width=0.9\linewidth]{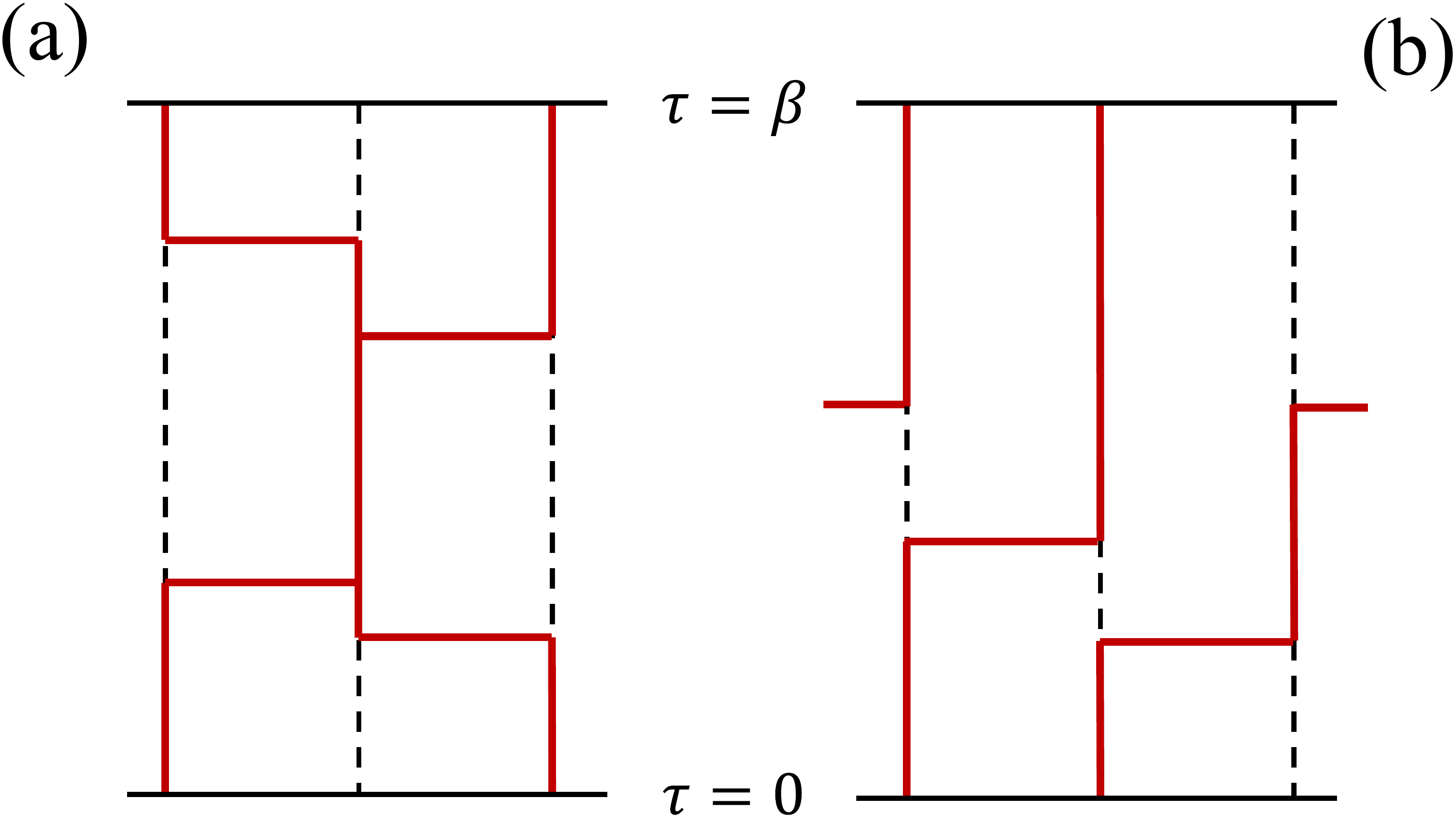}
    \caption{Schematic diagram of worldline configuration of a lattice model for (a) softcore particles and (b) hardcore particles. In panel (a), the solid line represents the path of softcore particles along the imaginary time direction, whose trajectories can intersect. Panel (b) shows a worldline configuration of hardcore particles where the paths of different particles cannot intersect. The worldlines of two particles connect at $\tau = \beta$, corresponding to a particle permutation, with $N_P = 1$ in this case.}
    \label{fig:worldline_configuration}
\end{figure}

Consider a system of $N$ particles with Hamiltonian $\mathcal{H}=\hat{K}+\hat{U}$, where $\hat{U}$ is diagonal in some expansion basis (potential part) and $\hat{K}$ is the off-diagonal part (kinetic part). Starting from the generalized partition function in Eq~\eqref{eq:generalized_partition_function}, one can perform standard Trotter decomposition and insert complete bases between every successive time slice~\cite{prokofev_worm_1998}. The path-integral formulation of the partition function is then given by,
\begin{align}
\mathcal{Z}(T, \xi) = \frac{1}{N!}\sum_{P \in S_N}\sum_{\mathcal{C}} \xi^{N_P} w(\mathcal{C}),
\end{align}
where the summation goes over all possible particle permutations and all worldline configurations $\mathcal{C}$, and $w(\mathcal{C})$ is the weight of the worldline configuration,
\begin{align}
w\left(\mathcal{C}\right) = K\left(\mathcal{C}\right) \exp \left[-U\left(\mathcal{C}\right)\right].
\label{eq:worldline_configuration_weight}
\end{align}
Here, $K\left(\mathcal{C}\right)$ is the weight factor due to off-diagonal terms, and $U\left(\mathcal{C}\right)$ is the total potential energy of the system. 
In the path‐integral representation, particle trajectories form closed loops in $(d+1)$‐dimensional space. The permutation $P$ associated with a worldline configuration can be identified by following the paths of individual particles from imaginary time $\tau=0$ to $\tau=\beta$. If the trajectory of one particle connects to the starting point of another at $\tau=\beta$, it indicates a pairwise exchange between the two particles, contributing a transposition to the overall permutation.

In conventional PIMC of lattice bosons, one typically adopts the occupation number basis $|n_1,n_2,n_3 \dots\rangle$ where $n_i$ is the number of particles on the $i$th site. Using this basis, particle exchange is implicit because the particles are indistinguishable by construction. By contrast, the fictitious‐particle PIMC requires a distinguishable‐particle basis, with quantum statistics imposed by explicitly sampling permutations. In this representation, each worldline corresponds uniquely to a specific particle, and the permutation $P$ of a given configuration is well-defined and directly measurable by tracing the worldline trajectory, as described above.

However, on a lattice, multiple worldlines may visit the same site simultaneously, making it difficult to unambiguously track individual trajectories, see Fig.~\ref{fig:worldline_configuration} (a). In continuous‐space PIMC, this issue is naturally avoided by imposing a hardcore potential that forbids two worldlines from occupying the same point. We adopt the same strategy in our lattice formulation by simulating hardcore particles whose worldlines never cross, as illustrated in Fig.~\ref{fig:worldline_configuration} (b). This choice is also well suited for studying the fermion limit $\xi\to-1$, where the Pauli exclusion principle prohibits two fermions from sharing a site. At $\xi = -1$, any intersection between two worldlines would lead to two configurations with equal weight but opposite sign, whose contributions cancel in the partition function. Therefore, the configuration space $\{\mathcal{C}\}$ remains identical for all values of $\xi$, which enhances the sampling efficiency of the algorithm.

Under the hardcore particle constraint, the total particle exchange number $N_P$ can be uniquely identified for each worldline configuration. This quantity arises naturally from the cycle decomposition of the underlying permutation $P$. In the worldline configuration, each distinct closed loop that winds along the imaginary time direction then corresponds to one cycle of $P$, and its winding number $k$ equals the cycle length. Mathematically, a cycle of length $k$ can be decomposed into $k-1$ pairwise transpositions. Therefore, the total exchange number of the worldline configuration is
\begin{equation}
  N_{P} = \sum_{\text{cycles}} (k-1) = N - N_{c},
  \label{eq:exchange-number}
\end{equation}
where $N$ is the total number of particles and $N_c$ is the number of closed loops in the worldline configuration.

\subsection{Worm Update Scheme for Fictitious Particles}
The worldline configuration of hardcore particles can be efficiently sampled using the worm algorithm, which has been highly successful in simulating various quantum and classical systems~\cite{prokofev_worm_1998,prokofev2001, boninsegni2006a, capogrosso-sansone2008,kornilovitch_path-integral_2007}. The worm algorithm operates in an enlarged configuration space by introducing an open-ended worldline known as a "worm." The worm's "head" and "tail" correspond to annihilation ($b$) and creation ($b^\dagger$) operators, respectively. Conventionally, the $b$-point is called \textit{Ira}, and the $b^\dagger$-point is called \textit{Masha}. The worldline configurations, consisting of closed particle trajectories, then form the $Z$ configuration space, while configurations that consist of worms form the $G$ space. Through local updates of \textit{Ira} and \textit{Masha} in the $G$ space, the algorithm efficiently samples the configuration with different winding numbers, and particle permutations (reconnecting two paths). The measurements are then performed when the configuration is back in $Z$ space. Despite being a local update scheme, the worm algorithm generally has a much smaller dynamical critical exponent than the Metropolis-type updates~\cite{prokofev_worm_1998, prokofev2001}; thus, it can be very efficient even near a phase transition.

A minimal worm algorithm of hardcore particles on a lattice system consists of the following updates:
\begin{enumerate}
    \item Create/Delete Worm: A worm is created on a randomly chosen segment (a graphical element along the imaginary axis where the state on a lattice site remains unchanged) or deleted when there is no obstacle between Ira and Masha.
    \item Move the worm head: Masha is moved along the imaginary time.
    \item Insert/delete a kink after a worm head: a kink is inserted or deleted at an imaginary time after the Masha, and the position of the Masha is changed
    \item Insert/delete a kink before a worm head: a kink is inserted or deleted at an imaginary time before the Masha, and the position of the Masha is changed
\end{enumerate}
This set of update operations ensures ergodicity and is capable of changing the winding number of the configuration and performing particle exchange.

In this work, we shall focus on the canonical ensemble where the total number of particles is fixed. The canonical ensemble is imposed by restricting the distance between two worm heads during the update: 
\begin{align}
|\tau_{\rm Ira} - \tau_{\rm Masha}| < \beta/2.
\end{align}
This constraint applies to create/delete worm and move worm head update. For the special case of create/delete worm update on a ring, we also reject any case that may change the particle number. Moreover, the hardcore limit is imposed by rejecting any proposed update that generates segments with occupation larger than $1$.

In the enlarged configuration space, a worldline configuration contains one open trajectory, i.e., the worm, which renders the previous definition of $N_P$ in the closed loop configuration invalid. To ensure the $N_P$ can be correctly tracked during the worm update, we extend the definitions of \(N_{c}\) and \(N_{P}\) in \(G\)-space so that they reduce to their closed‐loop values when the worm is removed. 

In the canonical ensemble, the particle number $N$ is fixed in $Z$-space. The worm algorithm temporarily violates this constraint, but only within a small window by restricting the separation between the worm head and tail to within $(-\beta/2, \beta/2)$. With this restriction, we can treat $N$ as effectively fixed in both $Z$-space and the worm space. We then generalize the definition of $N_c$, as the number of clusters in a worldline configuration. A cluster is a connected set of segments and kinks, so that both closed loops and the worm (open path) count as clusters. This definition is based on the fact that the worm will eventually be removed, forming a closed loop in $Z$-space. In this definition, there is a special case that requires careful handling when the length of the open trajectory is smaller than $\beta/2$. Such a trajectory does not represent a physical particle in the canonical ensemble, because if it is not connected to a longer loop by some kink update, all updates of this isolated worm will have no effect in the $Z$-space. 
To avoid counting such transient configurations in $G$-space, we only include clusters whose length is larger than $\beta/2$. 
Using the generalized definition of $N_P$, we can explicitly update and sample \(N_P \) along with worm moves, and incorporate generalized exchange statistics into the detailed balance condition.

\begin{figure}
    \centering
    \includegraphics[width=1.0\linewidth]{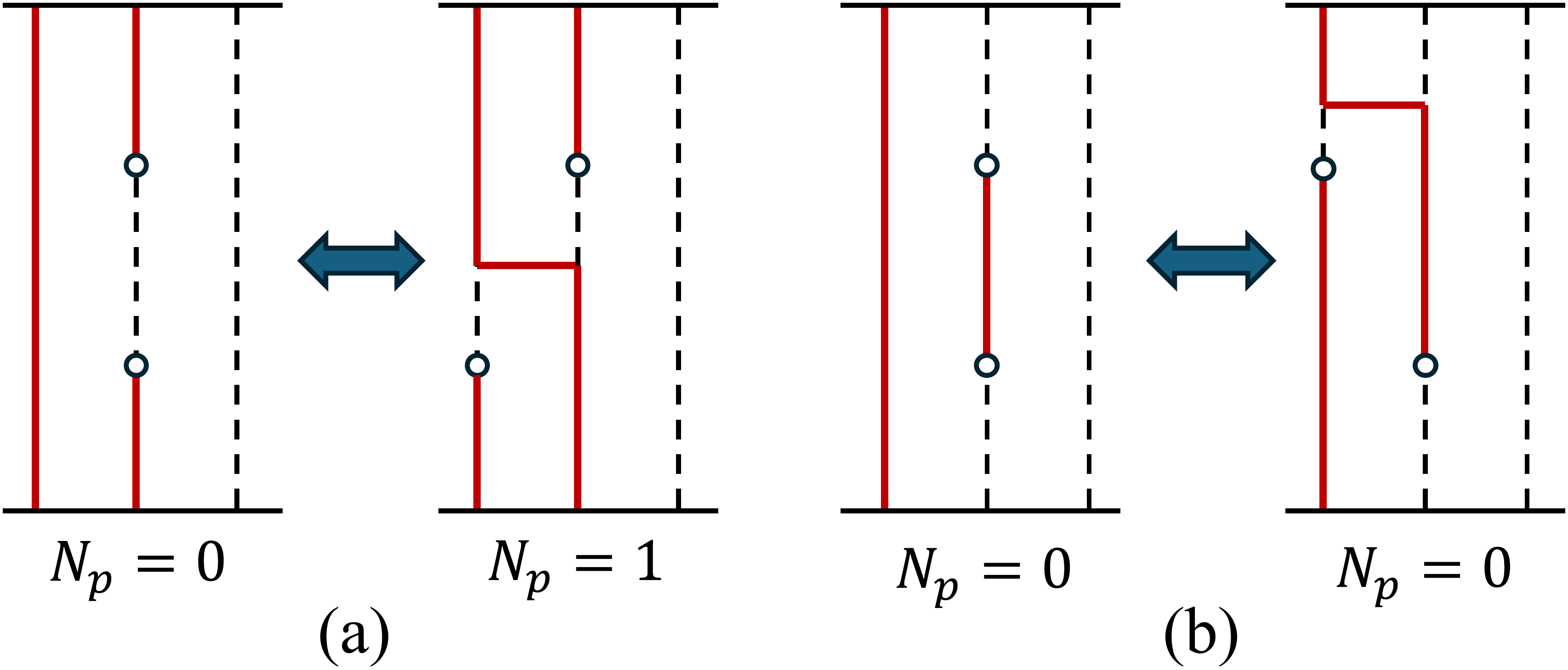}
    \caption{Insert/delete kink updates that (a) change $N_P$ and (b) do not change $N_P$. In panel (a), a worm is created by opening an existing worldline, resulting in two disconnected clusters. Inserting a kink that connects two different worldlines reduces the number of clusters $N_c$ by 1 and increases the total permutation number $N_P$ by 1. In panel (b), a worm is created by inserting a new worldline with length smaller than $\beta/2$, which is not associated with any physical particle. After kink insertion, $N_P$ remains unchanged, as this worm does not contribute to the physical permutation.}
    \label{fig:kink_update}
\end{figure}

In our PIMC algorithm for fictitious particles, only kink‐insertion and kink‐deletion moves can alter the total exchange number $N_{P}$, since they change the connectivity of worldline clusters—analogous to the swap update in continuous‐space PIMC~\cite{morresi_normal_2025}. The \textit{insert kink after worm head} update proposes a new creation–annihilation operator pair immediately after the worm head (Masha), allowing the particle to hop from its current site to a neighboring site. This corresponds to the insertion of a creation-annihilation operator pair, locally modifying the occupation configuration and the spatial trajectory of the worm. If, after the update, Masha connects to a previously disconnected worldline cluster, the number of disjoint cycles \(N_{c}\) decreases by one, and thus \(N_{P}=N-N_{c}\) increases by $1$, see Fig.~\ref{fig:kink_update} (a).  Conversely, if Masha hops onto the open trajectory itself, two cycles fuse and \(N_{P}\) decreases by one. For worm segments with temporal length shorter than \(\beta/2\), the worm is treated as virtual and does not contribute to \(N_{P}\), as shown in Fig.~\ref{fig:kink_update} (b). The reverse move removes the kink immediately following the worm head, restoring the original connectivity and accordingly adjusting \(N_{P}\) by \(\pm1\) as clusters split or merge.

The acceptance probability for a kink insertion or deletion then includes both energetic contributions and an additional statistical factor arising from particle exchange:
\begin{align}
p_{\text{accept}} = \min\left[1, \frac{p(\mathcal{C}\to \mathcal{C}') w(\mathcal{C}')}{p(\mathcal{C}'\to \mathcal{C}) w(\mathcal{C})}\xi^{\Delta N_P}\right],
\end{align}
where $\Delta N_P$ is the change in the particle exchange number due to the proposed update, and $p(\mathcal{C} \to \mathcal{C}')$ is the proposal probability of the update.

\subsection{Reweighting Measurement Technique}

The generalized partition function in Eq.\eqref{eq:generalized_partition_function} not only allows direct simulations of fictitious particles for $\xi>0$, but also enables a direct reweighting to $\xi<0$. The reweighting technique is a powerful numerical method used in Monte Carlo simulations to efficiently obtain physical observables across different parameter regimes without performing extensive independent simulations. One typically simulates the system at a given parameter and reweights the distribution to estimate the physical observables at different parameters near the original one.

The partition function of fictitious particles has the form
\begin{align}
    \mathcal{Z}\propto\sum_{\{\mathcal{C}\}} \xi^{N_P} w(\mathcal{C}).
\end{align} 
One can then apply the reweight method to $\xi$. Specifically, during a simulation performed at a certain $\xi_0$, we measure observables at another target $\xi$ by reweighting the measurement with the factor $(\xi/\xi_0)^{N_P}$. The expectation value of an observable $O$ at $\xi$ is then given by,
\begin{align}
    \langle O \rangle_{\xi}=\frac{\langle O \times (\xi/\xi_0)^{N_P}\rangle_{\xi_0}}{\langle(\xi/\xi_0)^{N_P}\rangle_{\xi_0}},
\end{align}
where $\langle \cdot \rangle_{\xi_0}$ is the ensemble average at $\xi_0$. Note that the reweighting should not be performed at $\xi_0 = 0$, where all configurations with $N_P \neq0$ are forbidden. 

In principle, the reweighting method allows us to evaluate physical observables at arbitrary values of $\xi$ using data obtained from a simulation performed at a fixed reference point $\xi_0 \neq0$. However, one should not reweight to a target parameter that is too far from $\xi_0$. For each reweighting target $\xi$ starting from a simulation at $\xi_0$, we measure the expectation value of $(\xi/\xi_0)^{N_P}$, which corresponds to the ratio of the partition functions at the two parameters:
\begin{align}
\langle \left(\xi/\xi_0\right)^{N_P} \rangle = \frac{Z_\xi}{Z_{\xi_0}}.
\end{align}
This quantity measures the average reweighting factor when transitioning from a simulation performed at $\xi_0$ to a target parameter $\xi$, thereby directly characterizing the statistical efficiency of the reweighting scheme. When the target $\xi<0$, we can interpret this ratio as a generalized sign, denoted as $s(\xi_0 \to \xi)$, which measures the magnitude of the sign problem of fictitious particle PIMC simulation. In particular, the conventional fermionic sign corresponds to $s(1 \to -1)$.

This reweighting framework provides an approach for probing quantum statistics and fermionic limits. When the generalized sign remains of order $\mathcal{O}(1)$, the sign problem is mild, and reweighting can be performed efficiently with well-controlled statistical errors. In practice, the reweighting approach is found to be rather effective in the intermediate-temperature regime or under strong interactions, where it enables direct estimation of observables in the fermionic limit.

\section{Hubbard Model of Fictitious Particles}
\label{section:hubbard_model}
In this work, we apply the fictitious particle PIMC method to the fictitious particle Hubbard model on a square lattice, a generalized Hubbard model of hardcore fictitious particles that interpolates between the two-component hardcore Bose-Hubbard model and the Fermi-Hubbard model. By studying this model, we can probe the properties of the Fermi-Hubbard model, which suffers from a severe sign problem in certain important parameter regimes.
The Hamiltonian of the fictitious particle Hubbard model is given by,
\begin{align}
    \mathcal{H}_\xi = -t\sum_{\langle i,j \rangle} (a^{\dagger}_{i\sigma} a_{j\sigma} + h.c.) + U \sum_i n_{i\uparrow} n_{i\downarrow},
\end{align}
where $t$ is the nearest-neighbor hopping strength and $U>0$ is the on-site repulsive interaction. Throughout this work, we set $t=1$ to define the unit of energy. The operator $a^{\dagger}_{i\sigma}$ ($a_{i\sigma}$) is the creation (annihilation) operator of a fictitious particle on $i$th site with (pseudo-)spin $\sigma = \uparrow,\downarrow$, and $n_{i\sigma}=a^{\dagger}_{i\sigma}a_{i\sigma}$ is the corresponding number operator. The creation and annihilation operators satisfy the generalized quantum exchange statistics in Eq.~\eqref{eq:generalized_exchange_statistics}. The sum in the hopping term is over all nearest neighbors. The model reduces to the Fermi-Hubbard model when $\xi=-1$, the model becomes the Fermi-Hubbard model, and to a two-component hardcore Bose-Hubbard model when $\xi=1$, which itself exhibits a variety of interesting phenomena~\cite{yanay2013,zeng2022}.

In this case, because the single-particle spin is a good quantum number, we treat each spin component as a separate species and sample permutations independently within each component~\cite{morresi_normal_2025}. The worldline configuration then has two components, one for each type of spin. The worldlines in different components do not intersect or connect to each other. The total permutation of a configuration can be factorized as $P = P_\uparrow P_\downarrow$, where $P_\uparrow$ is a permutation among particles with up spin, and $P_\downarrow$ is a permutation among particles with down spin. Thus, the total permutation number is $N_P=N_{P_\uparrow} + N_{P_\downarrow}$. We perform QMC simulation in the $\xi>0$ regime using the algorithm formulated in the previous section. During the update, the worm is created on a random component and updates only that spin component until the worm is deleted.

To characterize the system, we measure the following quantities for a given worldline configuration:
\begin{enumerate}
    \item The total energy
    \begin{align}
    E = \frac{ n_k }{\beta} + \frac{U}{\beta} \sum_i\int_0^\beta d\tau n_{i\uparrow}(\tau) n_{i\downarrow}(\tau),
    \end{align}
    where $n_{k}$ is the total number of kinks in the worldline configuration. The first term represents the kinetic contribution, and the second term represents the on-site interaction energy.
    \item The fraction of lattice sites occupied by two particles,
    \begin{align}
        d = \frac{1}{L^2} \sum_i \int_0^{\beta} d\tau n_{i\uparrow}(\tau)n_{i\downarrow}(\tau). 
    \end{align} 
    Here, $L$ is the linear size of the system.
    \item The staggered magnetization $M_{\mathrm{stagger}}$, which quantifies antiferromagnetic (AFM) spin correlations across the lattice, is defined as 
    \begin{align}
        M_{\mathrm{stagger}}= \frac{1}{L^2} \int_0^\beta  d\tau  \sum_i (-1)^{x_i+y_i} S^z_i (\tau),
    \end{align}
    where $S^z_i=\frac{1}{2}(n_{i\uparrow} - n_{i\downarrow})$ is the spin component in $z$-direction on the $i$th site and $x_i,y_i$ are the coordinates of the $i$th site.
\end{enumerate}
During the simulation, we measure the following physical observables,
\begin{enumerate}
    \item Energy density $\varepsilon=\langle E \rangle/L^2$.
    \item Double occupancy $\langle n_{i\uparrow} n_{i\downarrow} \rangle \equiv \langle d \rangle$.
    \item AFM structure factor $S(\pi,\pi) = L^2 \langle M^{2}_{\mathrm{stagger}} \rangle$.
\end{enumerate}
The $\langle \cdot \rangle$ represents the Monte Carlo average. 

To study the fermionic regime, we reweight the measured observables to various values of $\xi<0$. To monitor the sign problem and the efficiency of reweighting, we also measure the generalized sign, 
\begin{align}
    s(\xi_0 \to\xi) = \langle (\xi/\xi_0)^{N_P}\rangle.
\end{align}
This quantity characterizes the average statistical weight of configurations under reweighting from a reference value $\xi_0$ to a target $\xi$.

\subsection{Benchmark with Exact Diagonalization}
\begin{figure}
    \centering
    \includegraphics[width=1.0\linewidth]{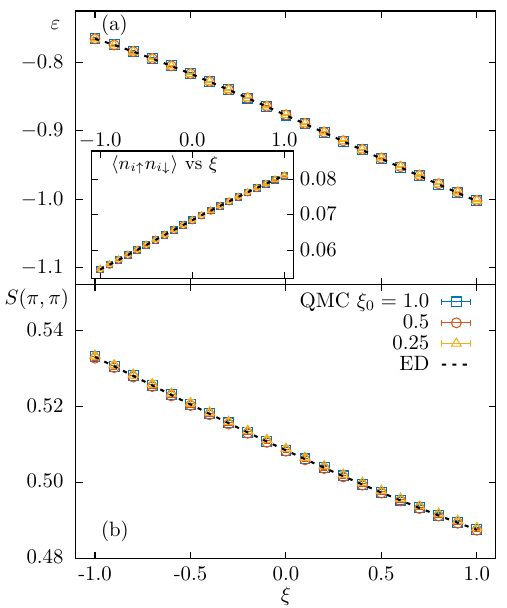}
    \caption{Comparison of QMC and ED results for (a) energy density, (inset) double occupancy, and (c) AFM structure factor $S(\pi,\pi)$ at half-filling on a $2 \times 2$ lattice with $U = 10$ and $\beta = 1$. The QMC results are obtained via reweighting from simulations at different $\xi_{0}$ values.}
    \label{fig:benchmark_results_half_filling_U_10}
\end{figure}

\begin{figure}
    \centering
    \includegraphics[width=1.0\linewidth]{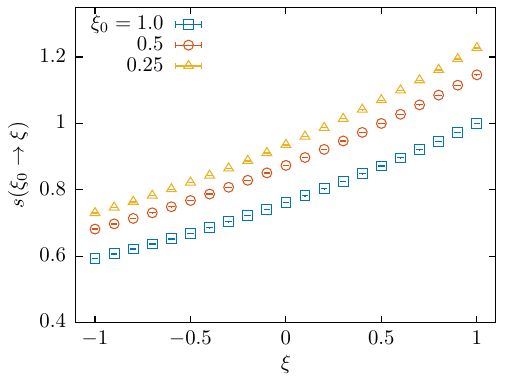}
    \caption{The generalized sign for simulations at $\xi_{0}=0.25,0.5,1.0$ at half-filling on a $2 \times 2$ lattice with $U = 10$ and $\beta = 1$.}
    \label{fig:benchmark_results_half_filling_U_10_ratio}
\end{figure}

As a benchmark, we compare the results of fictitious particle PIMC and the exact diagonalization (ED) method on a small system. 

To perform exact diagonalization for lattice models of fictitious particles, we begin by explicitly constructing the many-body Hilbert space in the real-space basis $|x_1,x_2,\dots x_N\rangle$. Each basis state represents a valid configuration of distinguishable quantum particles on the lattice, respecting the hardcore constraint. The total size of the Hilbert space of $N$ hardcore particles on $M$ sites is $M!/(M-N)!$, which is much larger than the conventional occupation basis for indistinguishable quantum particles. For particles with spin degrees of freedom, basis states are represented as ordered lists of occupied sites for each spin component $|x_{\uparrow,1},x_{\uparrow,2},\dots x_{\uparrow, N_\uparrow}; x_{\downarrow,1},x_{\downarrow,2},\dots x_{\downarrow, N_\downarrow}\rangle$. 

The generalized quantum statistics of particles is recovered by explicitly incorporating the full permutation group $S_N$. We first diagonalize the Hamiltonian matrix in the distinguishable particle basis. Then, for each eigenstate of the Hamiltonian $|\phi_j\rangle$, we compute its overlap with all permuted copies of itself, $\langle \phi_j | P(\phi_j) \rangle$, and evaluate the generalized partition function:
\begin{align}
    \mathcal{Z}_{\xi}(T) = \frac{1}{N!}\sum_j e^{-\beta E_j}\sum_{P\in S_N}\xi^{N_P}\langle \phi_j|P(\phi_j)\rangle.
\end{align}
This formalism then enables direct computation of thermodynamic observables for systems of fictitious particles with different $\xi$ values.

Figure~\ref{fig:benchmark_results_half_filling_U_10} compares the energy density, double occupancy, and AFM structure factor obtained from PIMC simulations with ED results. The calculations are performed on a $2 \times 2$ lattice with periodic boundary conditions at $U=10$ and $\beta=1$. The PIMC simulations are conducted at $\xi_0 = 0.25$, $ 0.5$, and $1.0$, while measurements at other $\xi$ are obtained by the reweighting method. The results for all $\xi_0$ values exhibit excellent agreement with the ED results across the entire range of $\xi$. As $\xi$ decreases from $1$ (bosons) to $-1$ (fermions), the energy density increases monotonically, while the double occupancy decreases monotonically. This behavior is consistent with the interpretation that quantum exchange statistics effectively induce repulsion, reducing the likelihood of multiple particles occupying the same site. Moreover, the $S(\pi,\pi)$ increases towards the fermionic limit, indicating an enhancement of short-range antiferromagnetic correlations as the system becomes more fermion-like.

Figure~\ref{fig:benchmark_results_half_filling_U_10_ratio} shows the generalized sign as a function of $\xi$ for different $\xi_0$. In the fermion limit $\xi=-1$, the average sign for simulation at $\xi_0 = 0.25$ is larger than that of the other $\xi_0$ values, i.e., the magnitude of the sign problem is smaller at $\xi_0 = 0.25$. This can be understood by noting that the factor $\xi^{N_P}$ in Eq.~\eqref{eq:generalized_partition_function} strongly suppresses the formation of permutation cycles as $\xi$ decreases~\cite{dornheim_fermionic_2023}, thereby reducing the cancellations between positive and negative weights. However, direct simulations near $\xi = 0$ may suffer from low acceptance rates and poor sampling efficiency, despite the improved behavior of the reweighting factor. This trade-off should be considered when choosing the optimal reference $\xi_0$ for reweighting.

\subsection{Half-filling System}
To evaluate the performance of the fictitious-particle PIMC approach, we first study the fictitious-particle Hubbard model at half-filling on larger lattices and across a broad range of parameters. In this regime, the interplay between quantum statistics and strong correlations gives rise to rich physical behavior.

\begin{figure}
    \centering
    \includegraphics[width=1.0\linewidth]{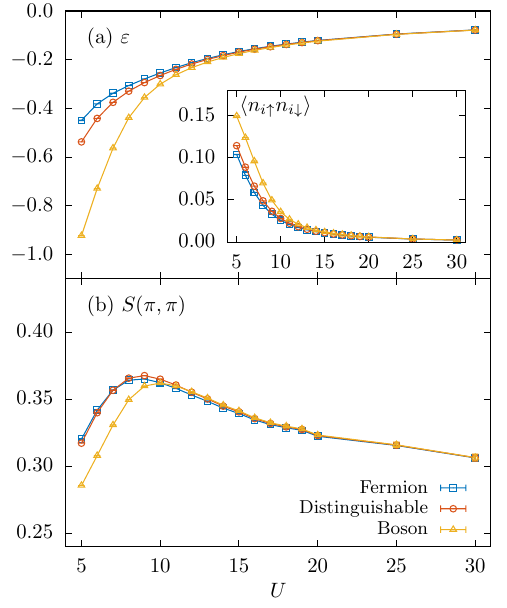}
    \caption{Various observables as functions of interaction strength $U$ for fermions, bosons, and distinguishable particles on a $4 \times 4$ lattice with $\beta=1$, obtained using the fictitious-particle PIMC. Panel (a) shows the energy density $\varepsilon$, with the inset displaying the double occupancy $\langle n_{i\uparrow} n_{i\downarrow} \rangle$. Panel (b) shows the AFM structure factor $S(\pi,\pi)$. The comparison reveals distinct behaviors among the three statistics, highlighting the role of quantum statistics in correlated many-body systems.}
    \label{fig:half_filling_obs_vs_U_various_limit}
\end{figure}

\begin{figure}
    \centering
    \includegraphics[trim={0 0.1cm 0 0},clip, width=1.0\linewidth]{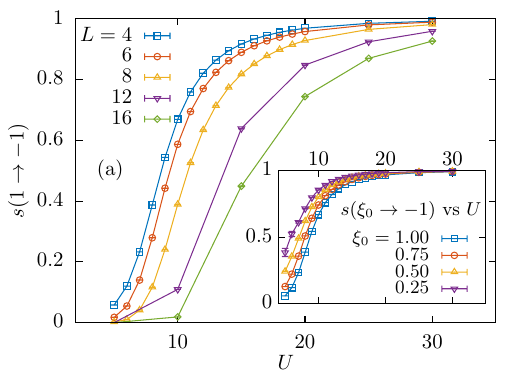}
    \includegraphics[trim={0.1cm 0 0 0},clip, width=1.0\linewidth]{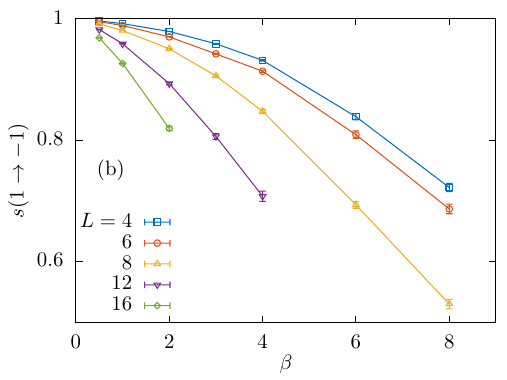}
    \caption{Average fermion sign of a half-filled system. Panel (a) shows the average sign as a function of $U$ at $\beta=1$ for various system sizes $L$, simulated at $\xi_0=1$. The sign problem is severe at small $U$ and becomes less pronounced as $U$ increases. The inset shows the average sign for $L=4$ and $\beta=1$ obtained from simulations with different $\xi_0$ values, indicating that simulations with smaller $\xi_0$ can yield a larger average sign. 
    Panel (b) shows the average sign as a function of $\beta$ at $U=30$ for various $L$, simulated directly at $\xi_0=1.0$. The sign problem of direct simulation gets more severe as $L$ and $\beta$ increase.
    }
    \label{fig:half_filling_average_sign}
\end{figure}

Figure~\ref{fig:half_filling_obs_vs_U_various_limit} shows the energy density, double occupancy, and structure factor $S(\pi,\pi)$ as functions of the interaction strength $U$ for a $4 \times 4$ lattice at inverse temperature $\beta=1$. Results are presented for three representative limits of quantum statistics: fermions ($\xi = -1$), bosons ($\xi = 1$), and distinguishable particles ($\xi = 0$). These data were obtained by averaging the reweighting results based on multiple simulations at different $\xi_0>0$. As will be explained later, the sign problem of the reweighting method is manageable within this parameter regime, and the reweighting method provides reliable estimates of observables across the full range of $\xi \in [-1,1]$.

For all three particle types, the double occupancy decreases rapidly when $U \gtrsim 10$, indicating the onset of a crossover from a weakly correlated metallic or superfluid regime to a Mott insulating state, where charge fluctuations are strongly suppressed due to the energetic cost of double occupancy. Notably, at fixed $U$, bosons exhibit the largest double occupancy, followed by distinguishable particles and then fermions, but the energy density of a bosonic system is lower than that in all other cases. These behaviors indicate that the wavefunction of particles becomes more localized, as $\xi\to-1$.

The structure factor $S(\pi,\pi)$ exhibits a broad peak around $U \approx 8$ for fermions and distinguishable particles, while for bosons the peak is located at $U \approx 10$. This enhancement indicates the development of short-range AFM correlation associated with the incipient Mott transition, even though true long-range AFM order is absent at the finite temperature and small system size considered here. For $U < 10$, fermions and distinguishable particles exhibit similar values of $S(\pi,\pi)$ while bosons have a much smaller $S(\pi,\pi)$. This suggests that the AFM correlations from configurations with $N_P > 0$ is weaker than those of $N_P=0$, and their contribution to total $S(\pi,\pi)$ is largely canceled in the fermion limit.
In contrast, at larger $U$, all observables become nearly independent of $\xi$, indicating that the effects of quantum statistics are strongly suppressed in the strongly interacting limit where charge localization dominates.

Figure~\ref{fig:half_filling_average_sign} (a) shows the average sign $s(1 \to -1)$ at half-filling as a function of interaction strength $U$ for various system sizes $L$ at $\beta = 1.0$. As the on-site repulsion $U$ increases, the average sign improves due to the suppression of particle exchanges, which reduces the occurrence of negative weight configurations and thus alleviates the sign problem. In contrast, the sign problem becomes more severe as the system size increases. This can be understood by assuming that particle exchange mostly occurs within a finite length scale; then, the average $N_P$ should increase as $\sim L^2$. As the average $N_P$ increases, the cancellation might become more severe. The inset of Fig.~\ref{fig:half_filling_average_sign} shows the average sign for $L=4$ and $\beta=1$ obtained by reweighting from different $\xi_0$ values. Simulations performed at smaller $\xi_0$ yield larger average signs.

Figure~\ref{fig:half_filling_average_sign} (b) further shows the average sign at $U=30$ as a function of $\beta$ for different system sizes with $\xi_0=1$. The observed decrease in the average sign with increasing $\beta$ reflects the fact that quantum exchange effects become more pronounced at lower temperatures, thereby enhancing the sign fluctuation of the configuration weight. Nonetheless, we expect the sign problem to remain manageable down to $\beta\approx12$ for $L=4$ and to $\beta\approx4$ for $L=16$. In such moderate system size and temperature, the system could exhibit interesting thermodynamic and dynamical properties~\cite{arovas2022}.

Our analysis suggests that the fictitious-particle PIMC method, combined with the $\xi$-reweighting technique, provides an effective approach for probing the Fermi-Hubbard model in the large-$U$ regime. In this regime, strong particle localization suppresses exchange processes, which in turn reduces the severity of the sign problem and facilitates access to fermionic observables via $\xi$-reweighting. On the other hand, while the DQMC method is sign-problem-free at half-filling, it often suffers from severe numerical instabilities and sampling issues at large $U$~\cite{khatami_finite-temperature_2015, assaad2008}. In contrast, the fictitious-particle PIMC method is expected to remain numerically stable and efficient even deep in the large-$U$ regime~\cite{capogrosso-sansone2008, sadoune2022}. Moreover, the PIMC method exhibits a more favorable time complexity of $\mathcal{O}(L^d\beta)$ compared to the $\mathcal{O}(L^{3d}\beta)$ scaling of DQMC~\cite{assaad2008}. These features suggest that the fictitious-particle PIMC method could serve as a valuable complement to conventional fermionic algorithms, particularly in the strong-coupling regime, where alternative approaches might become less stable or more computationally expensive.

\subsection{Doped System at Strong Coupling}

\begin{figure}[t]
    \centering
    \includegraphics[width=1.0\linewidth]{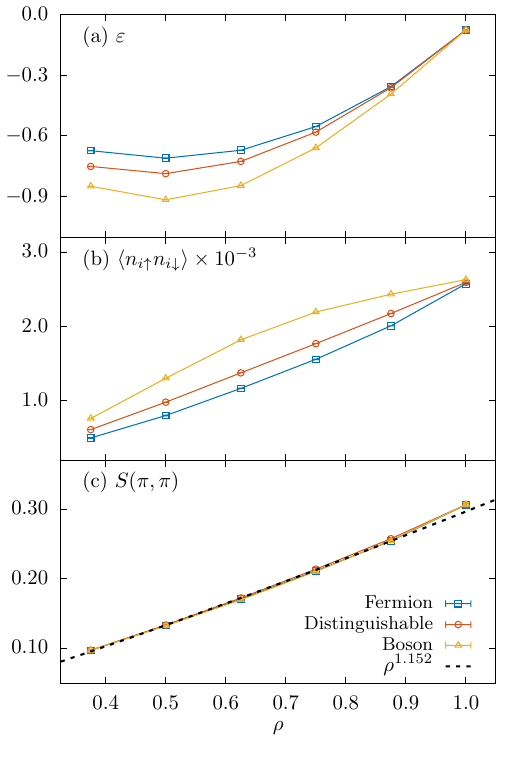}
    \caption{Various observables as functions of particle density $\rho$ for fermions, bosons, and distinguishable particles on a $4 \times 4$ lattice with $\beta=1$ and $U=30$: (a) energy density $\varepsilon$, (b) double occupancy $\langle n_{i\uparrow} n_{i\downarrow} \rangle$, and (c) structure factor $S(\pi,\pi)$. The system is unpolarized with equal populations of spin-up and spin-down particles, i.e., $N_{\uparrow} = N_{\downarrow}$.}
    \label{fig:obs_vs_density}
\end{figure}

\begin{figure}[t]
    \centering
    \includegraphics[width=1.0\linewidth]{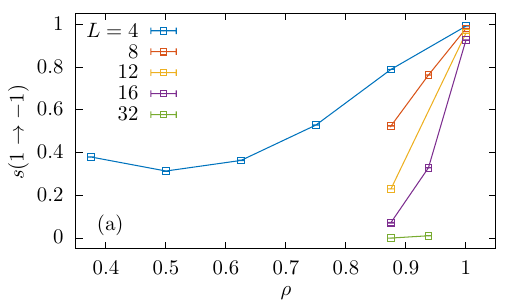}
    \includegraphics[width=1.0\linewidth]{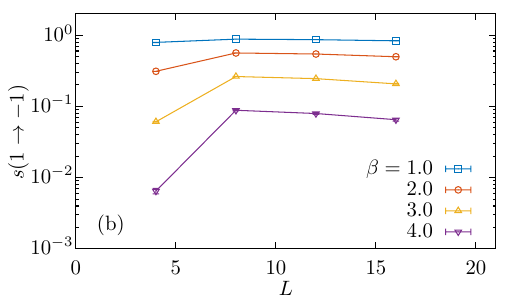}
    \includegraphics[width=1.0\linewidth]{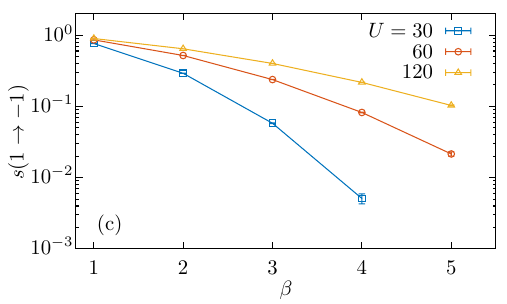}
    \caption{Average sign for the doped case. (a) Average sign versus density $\rho$ for various system sizes $L$ at $U=30$, $\beta=1.0$. (b) Average sign versus $L$ at $U=30$ for various $\beta$, with two hole dopants (one for each type of spin). (c) Average sign versus $\beta$ at $\rho=0.9375$ and $L=8$ for various interaction strengths $U$ (semi-log scale).}
    \label{fig:sign_doped}
\end{figure}

Based on our findings at half filling --- where $\xi$-reweighting becomes increasingly effective at large $U$ --- we further apply the fictitious-particle PIMC method to the hole-doped Hubbard model in the strong coupling regime, and examine the performance of the reweighting method. Upon doping, the exact mapping to the spin-$\frac{1}{2}$ Heisenberg model at large $U$ no longer holds, and the low-energy physics is instead approximately described by an effective $t$-$J$ model. This regime can host rich phenomena, such as Nagaoka polarons and kinetic ferromagnetism~\cite{arovas2022}, but also poses serious challenges for existing QMC methods.

Figure~\ref{fig:obs_vs_density} illustrates the energy density, the on‐site double occupancy, and the AFM staggered structure factor as functions of the total particle density $\rho$ on a $4\times4$ lattice at $\beta=1$ and $U=30$, for fermions ($\xi=-1$), distinguishable particles ($\xi=0$), and bosons ($\xi=1$). The particle density is defined as $\rho = (N_{\uparrow} + N_\downarrow)/L^2$. For simplicity, the system is kept unpolarized with equal populations of the two types of particles, i.e., $N_{\uparrow} = N_\downarrow$. As shown in panel (a), over the explored density range, bosons exhibit the lowest energy while fermions have the highest, with distinguishable particles in between. The energy density also shows a minimum at quarter filling $\rho=0.5$, where the energy difference among the three types of particles is most pronounced. Panel (b) shows the double occupancy (scaled by $10^{-3}$) as a function of $\rho$. It can be seen that bosons maintain the highest probability of double occupation, while fermions maintain the lowest. Another intriguing feature is that the curves are concave for bosons, approximately straight for distinguishable particles, and convex for fermions. Panel (c) reveals that $S(\pi,\pi)$ almost collapses onto a single curve for all three statistics, closely following a power‐law scaling $S(\pi,\pi)\propto\rho^{1.152}$ as shown by the dashed line. This indicates that, in the strong‐coupling regime $U\gg t$ of an unpolarized system, short‐range correlations are governed predominantly by the particle density and only weakly depend on the quantum exchange statistics.

Figure~\ref{fig:sign_doped} summarizes how the sign problem evolves with electron density and temperature for unpolarized doped systems at large $U$. In Fig.~\ref{fig:sign_doped} (a), we show the average sign $s(1\to-1)$ as a function of particle density $\rho$ for $L=4,8,16,32$ at $U=30$ and $\beta =1$. The curve for $L=4$ displays a pronounced ``U‐shape": the sign is close to $1$ at half filling, dips to a minimum around $\rho\approx0.5$, and then recovers as the system is emptied. This can be understood as follows. At half-filling, the sign oscillation is weak, since the particles are mostly localized due to the strong on-site repulsion $U$ and particle permutation is rare. In the dilute limit $\rho\to0$, the separation between particles is so large that particle exchange is unlikely to occur, which also leads to a weak sign oscillation. However, at the intermediate filling, particle exchanges become more frequent, causing a stronger sign cancellation. The U-shaped $\rho$ dependence of the sign in the fictitious particle PIMC differs from that of the DQMC at intermediate $U$, for which various dips occur between $0.5<\rho < 1.0$\cite{white1989}. Furthermore, for a fixed doping, the sign problem becomes more severe as $L$ increases. By $L=32$, the sign is approximately zero at $1/16$ hole doping, rendering the reweighting scheme inefficient. On the other hand, for a fixed number of dopants, the average sign decays much more slowly as the system size increases, as shown in Fig.~\ref{fig:sign_doped} (b). Such scaling is favorable when studying the physics in the limit of dilute holes. Figure~\ref{fig:sign_doped} (c) further shows the temperature dependence of the average sign for $L=8$ at $\rho=0.9375$ and various $U$. The sign decreases as $\beta$ increases. For $U=30$, it drops below $0.01$ at $\beta=4$, making direct reweighting inefficient at $\beta > 4$. For larger on-site repulsion $U$, however, the sign exhibits a slower decay. For instance, at $U=120$, the sign remains approximately $0.1$ down to $\beta = 5$. 

The overall behavior of the sign suggests that the $\xi$-reweighting method remains effective in the regime of small doping ($\delta \to 0$) and large interaction strength ($U \gg t$), corresponding to the so-called Nagaoka limit. It has been reported that, in this strongly correlated regime, the Hubbard model on a 2D square lattice exhibits Nagaoka polarons and short-range ferromagnetic correlations at finite temperature and large $U$~\cite{newby2025}. These observations underscore the potential of the fictitious-particle PIMC method, when combined with $\xi$-reweighting, as a powerful tool for investigating emergent phenomena in the doped, strongly interacting Hubbard model.

\section{Extrapolation to Fermion Limit}
\label{section:extrapolation}
In this section, we assess the performance of the direct $\xi$-extrapolation method for the 2D Hubbard model. The method leverages the continuity of observables like energy $E(T,\xi)$ to extrapolate from the sign-problem-free regime ($\xi>0$) to the fermionic limit ($\xi=-1$)~\cite{xiong2022}. While a simple parabolic ansatz for observables,
\begin{align}
O(T,\xi)=c_0 + c_1 \xi + c_2 \xi^2.
\label{eq:E_extrapolation}
\end{align}
has proven effective for the energy density in various interacting continuum systems~\cite{xiong_thermodynamic_2022, xiong2022} and for other observables such as correlation functions~\cite{dornheim_fermionic_2023}, its applicability and the required polynomial order are not guaranteed for lattice models, especially under strong correlations. Here, we test this approach at a strong coupling of $U=30$ and an intermediate inverse temperature of $\beta=1.0$ at $1/8$ hole doping.

\begin{figure}
    \centering
    \includegraphics[width=1.0\linewidth]{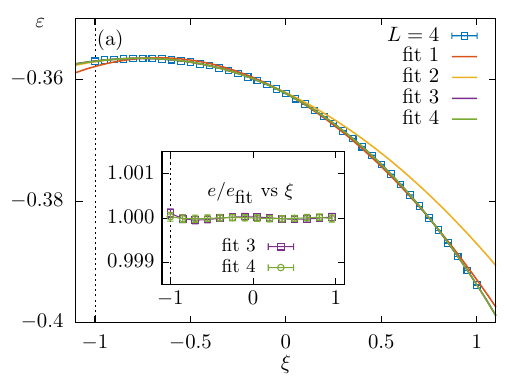}
    \includegraphics[width=1.0\linewidth]{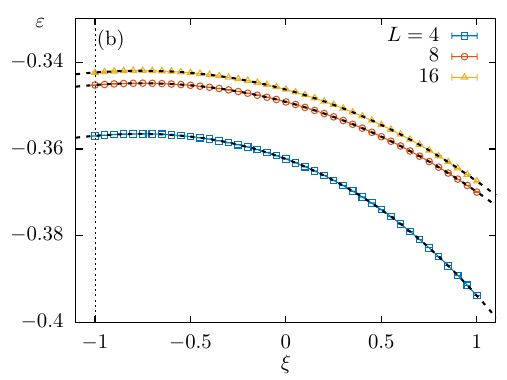}
    \caption{Energy density $\varepsilon$ as functions of $\xi$ at $U=30$, $\rho=0.875$, and $\beta=1.0$. Panel (a) compares different fitting ansatz and fitting ranges. Fit 1 stands for quadratic fit over $\xi\in[-1,1]$; fit 2 is the quadratic fit over $\xi\in[-1,0]$; fit 3 is the cubic fit over $\xi\in[-1,1]$; fit 4 is the quartic fit over $\xi\in[-1,1]$. The inset shows the ratio between the simulation data and the fitted function for fits 3 and 4. Panel (b) shows the $\varepsilon(\xi)$ curves for various system sizes $L=4,8,16$ can be well-fitted by cubic polynomials over the entire $\xi\in[-1,1]$ range.}
    \label{fig:extrapolation_of_energy_rho_0.875}
\end{figure}

We first applied the method to the energy density $\varepsilon$. We start by performing a series fit for the small system of $L=4$. In this case, the MC data over the entire $\xi$ range can be accurately obtained using the reweighting technique. As shown in Fig.~\ref{fig:extrapolation_of_energy_rho_0.875} (a), we compare the fit quality of various fitting ansatz and fitting ranges. The simple quadratic ansatz (fit 1) in Eq.\eqref{eq:E_extrapolation} does not provide a satisfactory fit to the full range of data for $\xi\in[-1,1]$. Both the fermionic limit and the bosonic limit show considerable discrepancy. We then change the fitting range by discarding data that is far from the $\xi=-1$ limit, and we find that only by dropping all $\xi>0$ data can the curve be well fitted using Eq.\eqref{eq:E_extrapolation}, with $\chi^2/\text{d.o.f.} \approx 1.3$ (fit 2). As suggested in Ref.~\cite{xiong2022}, we also gradually increase the order of the ansatz. We find that a third-order polynomial indeed yields a substantially improved fit over the entire $\xi$ range (fit 3), and a quartic polynomial also provides an excellent fit (fit 4). The ratios between the MC data and the fit curves for fits 3 and 4 are shown in the inset. The quartic fit, however, does not provide a significant qualitative improvement over the cubic fit, suggesting that a third-order polynomial suffices in this case as the extrapolation ansatz. We then fit the \(L=8\) and \(L=16\) data using the cubic polynomial, observing similarly high-quality fits, as shown in Fig.~\ref{fig:extrapolation_of_energy_rho_0.875} (b). This result indicates that, under the given conditions, a cubic polynomial is required to accurately describe the functional dependence of $\varepsilon(\xi)$ in the Hubbard model.

\begin{figure}
    \centering
    \includegraphics[width=1.0\linewidth]{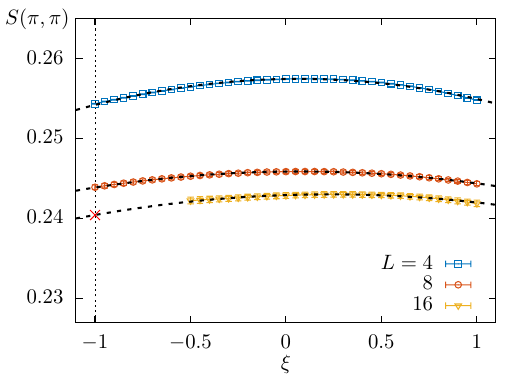}
    \caption{The AFM structure factor $S(\pi,\pi)$ as functions of $\xi$ at $U=30$, $\rho=0.875$, and $\beta=1.0$ for system sizes $L=4,8,16$. For $S(\pi,\pi)$, the data for $L=4$ and $L=8$ can be fitted by a quadratic polynomial over the entire range of $\xi$. Measurements of $S(\pi,\pi)$ at $\xi < -0.5$ show large statistical uncertainties and are thus not shown. By fitting the statistically reliable regime with a quadratic polynomial, we extrapolate the $S(\pi,\pi)$ to the $\xi=-1$ limit, marked by the red cross.}
    \label{fig:extrapolation_of_s_pi_pi_rho_0.875}
\end{figure}

Next, we examine the AFM structure factor $S(\pi,\pi)$ as a function of $\xi$ as shown in Fig.~\ref{fig:extrapolation_of_s_pi_pi_rho_0.875}. It can be seen that distinguishable particles have the highest $S(\pi,\pi)$, with the fermionic value slightly below the bosonic one. This indicates that even though the quantum exchange effect is suppressed at strong coupling, the presence of holes induces particle exchanges and reduces the overall AFM correlation. However, this observable is considerably more sensitive to the fermion sign problem. In particular, for \(L=16\) the relative error in \(S(\pi,\pi)\) becomes prohibitively large for \(\xi<-0.5\), rendering direct reweighting unreliable in this region. To address this limitation, we adopted a systematic extrapolation strategy. For smaller systems with $L=4,8$, where the sign problem is less severe, we find that a quadratic polynomial provides an excellent fit for $S(\pi,\pi)$ across the entire $\xi$ range. Based on this observation, we applied the same quadratic ansatz to the $L=16$ data, restricting the fit to the statistically reliable window $\xi \in [-0.5, 1.0]$ as shown in Fig.\ref{fig:extrapolation_of_s_pi_pi_rho_0.875}. Extrapolating this fit to $\xi = -1$, we estimate that $S(\pi,\pi)\big|_{\xi=-1}\approx0.2404$ in the fermion limit. Currently, no reliable benchmark data exist in this parameter regime for direct comparison, highlighting the importance of such extrapolations for exploring strongly interacting, doped fermion systems.

Our analysis demonstrates that $\xi$-extrapolation, when combined with reweighting, provides a viable approach for probing the Fermi-Hubbard model. By incorporating data from both the sign-problem-free regime ($\xi \ge 0$) and the fermionic sector ($\xi < 0$) obtained via reweighting, we achieve a more controlled and reliable extrapolation.
The reliability of the extrapolation also depends on the observable considered and the choice of fitting ansatz. At the considered temperature, we find that a low-order polynomial (quadratic or cubic) suffices for accurately capturing the $\xi$-dependence of observables such as energy density and spin correlations. This suggests that, in this regime, the observable curves remain sufficiently smooth to enable direct extrapolation. At lower temperatures or in more complex parameter regimes, advanced extrapolation schemes~\cite{xiong2023, morresi_normal_2025} may be necessary. In particular, the Appendix provides a discussion of the low-temperature properties of the fictitious-particle partition function, which may influence the applicability of extrapolation schemes.

\section{Discussion and Conclusion}
\label{section:discussion}

In this work, we demonstrate that the fictitious particle PIMC method, a seemingly brute-force approach, is surprisingly effective for studying the two-dimensional Hubbard model at strong coupling. As the interaction strength $U$ increases, the fermion sign problem becomes less severe due to suppressed particle exchange, allowing direct $\xi$-reweighting to remain viable even in doped systems at moderate temperatures. For parameter regimes where the sign problem is severe, we show that combining reweighting with $\xi$-extrapolation enables a more controlled and reliable extrapolation to the fermion limit. We reveal that $\xi$-extrapolation and low-order polynomial ansatz suffice for controlled extrapolation to the fermion limit. For $U=30$, $\beta =1$, $\rho=0.875$, a third‐order ansatz accurately describes the energy across the full range of $\xi$, while a quadratic fit captures the AFM structure factor. These results establish the validity of the $\xi$-extrapolation method for lattice fermions and extend the applicability of the fictitious particle framework.
Overall, the combined use of fictitious-particle PIMC, reweighting, and $\xi$-extrapolation offers a versatile and complementary toolset for investigating strongly correlated lattice systems. This approach offers a promising route to explore parameter regimes that challenge conventional fermionic algorithms, such as the Nagaoka limit~\cite{nagaoka1966, arovas2022}, and deepen our understanding of spin–charge interplay in doped Mott insulators.

We also note that several difficulties remain in studying the lattice model using the fictitious particle PIMC method. At large $U$, though the sign problem is small due to suppressed particle exchange, the efficiency of the worm update can be reduced. The worldline configuration of virtual exchange of spins can be very difficult to sample using the standard worm update scheme. A small segment with double occupancy is severely penalized by an exponential factor of $U$. To overcome these sampling bottlenecks, advanced techniques such as directed-loop algorithms or cluster-style moves are promising next steps~\cite{sadoune2022}. On the other hand, at lower temperatures, both reweighting and $\xi$-extrapolation become increasingly delicate, suggesting that further tempered extrapolation strategies will be necessary to access this regime. Searching for strategies to expand the accessible parameter regime and validating them constitutes the current frontier of fictitious-particle-based Monte Carlo methods. Further possible improvement may involve incorporating other tools for fermion PIMC simulations to mitigate the cancellations due to particle exchange, such as the permutation blocking paradigm~\cite{dornheim2015}.

The underlying formalism of the fictitious particle PIMC is also highly versatile. A straightforward generalization to the grand-canonical ensemble would permit fluctuations in particle number, enabling us to access experimentally relevant systems. Conversely, a micro-canonical treatment, where particle exchange numbers are kept fixed, could help dissect the structure of the sign problem itself. In addition, the fictitious-particle PIMC provides a natural method for studying the effect of particle distinguishability~\cite{boninsegni2012,menssen2017}, which is separated from other quantum effects in the path-integral formulation, as a function of the parameter $\xi$. This can provide novel insights for the emergence of exotic quantum phases in quantum many-body systems.

\acknowledgments
We thank Yunuo Xiong, Hongwei Xiong, and Tommaso Morresi for useful discussions. We acknowledge the support by the National Natural Science Foundation of China (NSFC) under Grant No. 12204173 and No. 12275263, as well as the Innovation Program for Quantum Science and Technology (under Grant No. 2021ZD0301900). YD is also supported by the Natural Science Foundation of Fujian Province 802 of China (Grant No. 2023J02032).
\\
\appendix

\section*{Appendix A: Low-temperature properties of Fictitious Particle Partition Function}
\label{appendix_a}

To better understand the limitations of the $\xi$-extrapolation method, we analyze the properties of the fictitious-particle partition function. Here, our analysis assumes that the Hamiltonian is fully symmetric under the permutation group $S_N$~\cite{lieb2009}, so that all permutation sectors ($N_P = 0,1,\dots,N-1$) can, in principle, contribute to the sum.

We start by reformulating the partition function to resolve contributions from different eigenstates:
\begin{align}
    \mathcal{Z}_{\xi}(T) &= \frac{1}{N!}\sum_{P \in S_N} \sum_{i} \xi^{N_P}\left\langle \Psi_i \right|e^{-\beta \mathcal{H}}\left| P(\Psi_i) \right\rangle\\
    &=\frac{1}{N!}\sum_{P \in S_N} \sum_{i} \xi^{N_P} \sum_j\left\langle \Psi_i \right| j \rangle e^{-\beta \varepsilon_j} \langle j \left| P(\Psi_i) \right\rangle\\
    &=\frac{1}{N!} \sum_j e^{-\beta \varepsilon_j} \sum_{P \in S_N} \xi^{N_P} \sum_{i} \langle j \left| P(\Psi_i) \right\rangle \left\langle \Psi_i \right| j \rangle 
\end{align}
where $|j\rangle$ is the eigenstate of $\mathcal{H}$ with energy $\varepsilon_j$. We define the self-overlap factor $a(j, P) \equiv \sum_{i} \langle j \left| P(\Psi_i) \right\rangle \left\langle \Psi_i \right| j \rangle$, which measures the projection of $|j\rangle$ onto its permuted representation in the chosen basis. The sum $\sum_i \left| P(\Psi_i) \right\rangle \left\langle \Psi_i \right|$ is the projection matrix of permutation $P$, which projects a wavefunction from the original basis to a permuted basis. The partition function can then be expressed as:
\begin{align}
    \mathcal{Z}_{\xi}(T) &=\frac{1}{N!} \sum_j e^{-\beta \varepsilon_j} \sum_{P \in S_N} \xi^{N_P}a(j,P)\\
    &=\sum_j e^{-\beta \varepsilon_j} \lambda_j(\xi),
\end{align}
where 
\begin{align}
    \lambda_j(\xi) =\frac{1}{N!}  \sum_{P \in S_N} \xi^{N_P}a(j, P)
\end{align}
is a polynomial of degree $N-1$ in $\xi$ that encodes the symmetry of $|j\rangle$ under $S_N$. It controls the weight of the eigenstate in the partition function. For example, a fully bosonic eigenstate satisfies $\lambda_j(1)=1$ and $\lambda_j(-1)=0$, so its contribution vanishes in the fermionic limit.

By further grouping permutations by $N_P$, we note that the generalized partition function can be written as an $(N-1)$th-degree polynomial in $\xi$:
\begin{align}
    \mathcal{Z}_{\xi}(T) &= \sum_{s=0}^{N-1} c(s,\beta) \xi^{s},
\end{align}
where $c(s,\beta)= \frac{1}{N!}\sum_{\{P|N_P = s\}} \sum_j e^{-\beta \varepsilon_j} a(j,P)$ collects the contributions from all permutations with $N_P=s$. One would then expect the function to have $N-1$ roots in the complex plane of $\xi$, which correspond to Lee-Yang zeros~\cite{yang1952, lee1952}. If the nearest zeros lie close to the real axis, they may induce non-analytic behavior in observables $O(\xi)$, hindering stable direct $\xi$-extrapolation.

In the limit $T \to 0$, the $\xi$-dependence of partition function is dominated by the ground state. We analyze the partition function of the 1D and 2D non-interacting lattice fermions with or without a hardcore constraint using exact diagonalization. The numerical analysis reveals that the partition function at low temperatures seems to have roots on the negative real axis at $-1,-\frac{1}{2},-\frac{1}{3},\cdots,-\frac{1}{N-1}$. This phenomenon can be understood as follows. For a fully permutation-symmetric Hamiltonian, the unrestricted ground state is bosonic~\cite{lieb2009}, implying $a(0,P) = 1$ for all $P$. The coefficients of $\lambda_0(\xi)$ polynomial are then given by the number of cyclic decompositions of the permutation group, which is related to the first Stirling number $s(N, N-N_P)$. The minimum number of pairwise swaps equals $N-N_c$, where $N_c$ is the number of cycles in the permutation. The generator of unsigned Stirling numbers of the first kind is
\begin{align}
    x(x+1)(x+2)...(x+N-1) = \sum_{N_c=0}^{N-1}|s(N,N_c)|x^{N_c}.
\end{align}
We then have $|s(N, N-N_P)|$ permutations with a given $N_P$. In the $T\rightarrow0$ limit, the partition function can be written as:
\begin{align}
    \mathcal{Z}_{\xi}(T) &\approx\frac{1}{N!} e^{-\beta \varepsilon_0}\sum_{N_p=0}^{N-1}|s(N,N-N_{P})|\xi^{N_P} \\
     &=  \frac{1}{N!}e^{-\beta \varepsilon_0} \xi^{N}\sum_{N_p=0}^{N-1}|s(N,N-N_{P})|\xi^{N_P-N}\\
     &= \frac{1}{N!}e^{-\beta \varepsilon_0} (1+\xi)(1+2\xi)...(1+(N-1)\xi)\\
     &= \frac{1}{N} e^{-\beta \varepsilon_0}(\xi+1)(\xi+\frac{1}{2})...(\xi+\frac{1}{N-1}).
\end{align}
Thus, for $N \ge 2$, the partition function has roots at $\xi = -1, -\frac12, -\frac13, \dots, -\frac{1}{N-1}$ in the $T\to0$ limit (for a fully symmetric Hamiltonian). The presence of these roots could therefore hinder extrapolation along the real $\xi$ axis at low temperatures.

At higher temperatures, excited-state contributions generally shift the zeros away from the real axis into the complex plane. In practice, once the nearest zeros are sufficiently far from the real axis compared to the scale set by statistical uncertainties, direct $\xi$-extrapolation becomes feasible. The quantitative threshold for “sufficiently far” will depend on the observable, noise level, and chosen fit ansatz. A more detailed investigation into the evolution of Lee–Yang zeros is ongoing and lies beyond the scope of the present work.

The Lee–Yang zeros of $\mathcal{Z}_\xi$ thus provide valuable insight into the analytic structure of the fictitious-particle method and may help guide the design of improved extrapolation strategies. We note that a recent independent work~\cite{he2025} provided a mathematical proof of the zero distribution at $T=0$ using a recursion relation for the canonical partition function, obtaining results consistent with our low-temperature analysis.

\bibliography{reference_v2}

\end{document}